\documentclass[a4paper,12pt]{article}
\usepackage[english]{babel}
\usepackage{graphicx,color}
\usepackage{amsmath,amsthm,amssymb,amsfonts}
\usepackage{exscale}
\usepackage{ifthen}
\usepackage{epsfig}
\usepackage[T1]{fontenc}
\usepackage[cp1250]{inputenc}
\usepackage[bookmarks,colorlinks]{hyperref}
\usepackage{fancyhdr}
\fancypagestyle{plain}{%
\fancyhead{}

}

\hypersetup{colorlinks,%
linkcolor=magenta,%
urlcolor=blue,%
citecolor=red}

\title{Energy renormalization and integrability\\within the massive neutrinos model}

\author{
\textbf{\L ukasz Andrzej Glinka}\vspace*{15pt}\\
E-mail: \href{mailto:laglinka@gmail.com}{\bf{\tt{laglinka@gmail.com}}}
\vspace*{10pt}\\
\emph{International Institute for Applicable}\\
\emph{Mathematics \& Information Sciences,}\\
\emph{Hyderabad (India) \& Udine (Italy),}\vspace*{10pt}\\
\emph{B.M. Birla Science Centre,}\\
\emph{Adarsh Nagar, 500 063 Hyderabad, India}
}
\date{\today}

\begin{document}

\maketitle
\begin{abstract}
In this paper the massive neutrinos model arising due to the Snyder noncommutative geometry, proposed recently by the author is partially developed. By straightforward calculation it is shown that the masses of the chiral left- and right-handed Weyl fields treated as parameters fixed by experiments, lead to the consistent physical picture of the noncommutative geometry, and consequently yield renormalization of an energy of a relativistic particle and exact integrability within the proposed model. This feature of the model in itself both defines and emphasizes its significance and possible usefulness for both theory as well as phenomenology for high energy physics and astrophysics.\\

\noindent \textbf{Keywords} models of neutrino mass ; noncommutative geometry ; Snyder model ; energy renormalization ; exactly integrable systems ; Planck scale effects\\

\noindent \textbf{PACS} 14.60.St; 03.65.Pm ; 02.40.Gh ; 03.30.+p
\end{abstract}
\newpage

\section{Introduction}
It was established recently by the author that the Dirac equation modified by the $\gamma^5$-term arising due to the Snyder noncommutative geometry model, yields the conventional Dirac theory with nonhermitian mass, or equivalently to the massive neutrinos model given by the Weyl equation with a diagonal and hermitian mass matrix. The model describes 4 massive chiral fields related to any original, \emph{i.e.} non-modified, massive or massless quantum state. Due to spontaneous global chiral symmetry breaking mechanism it leads to the isospin-symmetric effective field theory, that is composed chiral condensate of massive neutrinos. All these results violate CP symmetry manifestly, however, their possible physical application can be considered in a diverse way. On the one hand the effective theory is beyond the Standard Model, yet can be considered as its part due to the noncommutative geometry model contribution. On the other in the massive neutrinos model masses of the two left- and two right-handed chiral Weyl fields arise due to mass and energy of an original state, and a minimal scale (\emph{e.g.} the Planck scale), and its quantum mechanical face becomes the mystic riddle.

This paper is mostly concentrated on the quantum mechanics aspect. It is shown that the model in itself yields consistent physical explanation of the Snyder noncommutative geometry model and consequently leads to energy renormalization of an original quantum relativistic particle. Computations arising directly from the Schr\"odinger formulation of both the Dirac and the Weyl equations, are presented. First, the manifestly non hermitian modified Dirac Hamiltonian is discussed. Its integrability is formulated by straightforward application of the Zassenhaus formula for exponentialization of sum of two noncommuting operators. It is shown, however, that this approach does not lead to well-defined solutions; for this case the exponents are still sums of two noncommuting operators, so that this procedure has a cyclic problem which can not be finished, and by this reason is not algorithm. For solving the problem instead of the Dirac equation we employ the Weyl equation with pure hermitian mass matrix rewritten in the Schr\"odinger form. Its integration is straightforward and elementary. We present computations in both the Dirac and the Weyl representations of the Dirac gamma matrices.

The paper is organized as follows. The Section 2 presents the motivation for further studies - the massive neutrinos model is recalled briefly. Next, in the Sections 3 particle's energy renormalization is discussed. In the Section 4 we present integrability problem for the modified Dirac equation. Section 5 is devoted for the massive Weyl equation integrability, and the Section 6 discusses some special case related to ultra-high energy physics. Finally in the Section 7 the results of the entire paper are summarized briefly.

\section{The massive neutrinos model}
Let us recall briefly the massive neutrinos model resulting from \cite{glinka}. The starting point is the noncommutative geometry model \cite{ng} of phase-space and space of a relativistic particle due to a fundamental scale $\ell$ proposed by Snyder \cite{snyder} (Cf. also Ref. \cite{battisti}), and given by the following lattice model \cite{lattice}
\begin{equation}\label{lattice}
x=ndx\quad,\quad d x=\ell\quad,\quad n\in\mathbb{Z}\quad\longrightarrow\quad\ell=\dfrac{l_0}{n}e^{1/n}\quad,\quad\lim_{n\rightarrow\infty}\ell=0,
\end{equation}
where $l_0>0$ is a constant, together with the De Broglie formula relating the coordinate $x$ with its conjugate momentum $p$
\begin{equation}\label{debroglie}
  p=\dfrac{\hbar}{x}.
\end{equation}
Application of the Kontsevich star-product \cite{kontsevich} to the phase space $(x,p)$ and two space points $x$ and $y$
\begin{eqnarray}
x\star p&=&px+\sum_{n=1}^\infty \left(\dfrac{\alpha i\hbar}{2}\right)^nC_n(x,p),\label{star1}\\
x\star y&=&xy+\sum_{n=1}^\infty \left(\dfrac{i\beta}{2}\right)^nC_n(x,y),\label{star2}
\end{eqnarray}
where for correctness $\alpha\sim 1$, $\beta$ are dimensionless constants, and $C_n(f,g)$ are the Hochschild cochains, is yielding to the deformed Lie brackets (For review of deformation quantization see \emph{e.g.} Ref. \cite{dito})
\begin{eqnarray}
\left[x,p\right]_\star&=&\left[x,p\right]+\sum_{n=1}^\infty \left(\dfrac{\alpha i\hbar}{2}\right)^nB_n(x,p),\label{starb1}\\
\left[x,y\right]_\star&=&\left[x,y\right]+\sum_{n=1}^\infty \left(\dfrac{i\beta}{2}\right)^nB_n(x,y),\label{starb2}
\end{eqnarray}
where $B_n(f,g)\equiv C_n(f,g)-C_n(g,f)$ are the Chevalley cochains. Using $[x,p]=-i\hbar$ and $[x,y]=0$, and doing the first approximation one obtains
\begin{eqnarray}
\left[x,p\right]_\star=-i\hbar+\dfrac{\alpha i\hbar}{2}B_1(x,p)\quad,\quad\left[x,y\right]_\star=\dfrac{i\beta}{2}B_1(x,y).\label{starb1a}
\end{eqnarray}
or in the Dirac ''method of classical analogy'' form \cite{dirac}
\begin{eqnarray}
\dfrac{1}{i\hbar}\left[p,x\right]_\star=1-\dfrac{\alpha}{2}B_1(x,p)\quad,\quad\dfrac{1}{i\hbar}\left[x,y\right]_\star=\dfrac{\beta}{2\hbar}B_1(x,y).\label{starb2a}
\end{eqnarray}
Because, for any $f,g\in C^\infty(M)$ holds $B_1(f,g)=2\theta(df\wedge dg)$, one obtains
\begin{eqnarray}
\dfrac{1}{i\hbar}\left[p,x\right]_\star=1-\dfrac{\alpha}{\hbar}(dx\wedge dp)\quad,\quad\dfrac{1}{i\hbar}\left[x,y\right]_\star=\dfrac{\beta}{\hbar} dx\wedge dy,\label{starb2b}
\end{eqnarray}
where $\hbar$ in first relation was introduced for dimensional correctness. Applying now the lattice model (\ref{lattice}) and the De Broglie relation (\ref{debroglie}) one receives
\begin{eqnarray}
\dfrac{i}{\hbar}\left[x,p\right]_\star=1+\dfrac{\alpha}{\hbar^2}\ell^2p^2\quad,\quad\dfrac{i}{\hbar}\left[x,y\right]_\star=-\dfrac{\beta}{\hbar}\ell^2,\label{nd1}
\end{eqnarray}
that defines the Snyder model. Note that this model was studied in some aspect by previous authors \cite{kadyshevsky}, but the model is related to this direction to a slight degree. The model developed in this paper arise mostly from the idea of the papers \cite{glinka}.

If we consider $\ell$ as a minimal scale, \emph{e.g.} Planck or Compton scale, then the model (\ref{nd1}) can be rewritten in terms of the maximal energy $\epsilon$
\begin{equation}\label{nd}
\dfrac{i}{\hbar}[x,p]=1+\dfrac{1}{\epsilon^2}c^2p^2\quad,\quad \dfrac{i}{\hbar}[x,y]=O\left(\dfrac{1}{\epsilon^2}\right)\quad,\quad \epsilon\equiv\dfrac{\hbar c}{\sqrt{\alpha}\ell}.
\end{equation}
The lattice model (\ref{nd}) straightforwardly yield the contribution to the Einstein Hamiltonian constraint of Special Relativity
\begin{equation}\label{ss1}
  E^2-c^2p^2\equiv(\gamma^\mu p_{\mu})^2=m^2c^4+\dfrac{1}{\epsilon^2}c^4p^4\quad,\quad p_{\mu}=[E,cp],
\end{equation}
where $m$ and $E$ are mass and energy of a particle, and consequently leads to the generalized Sidharth $\gamma^5$-term within the usual Dirac equation
\begin{equation}\label{dsi}
  \left(\gamma^\mu \hat{p}_\mu\pm mc^2\pm \dfrac{1}{\epsilon}c^2\hat{p}^2\gamma^5\right)\psi=0\quad,\quad \hat{p}_\mu=i\hbar[\partial_0,c\partial_i],
\end{equation}
violating the Lorentz symmetry manifestly. In fact the equation (\ref{dsi}) describes 4 cases, that are dependent on the choice of the signs of mass $m$ and the $\gamma^5$-term. Here we will consider, however, positive signs case only. The negative ones are due to the changes $\epsilon\rightarrow-\epsilon$ and $m\rightarrow -m$ in the results obtained from the positive signs case.

Preservation of the Minkowski momentum space structure within the modified Einstein constraint (\ref{ss1})
\begin{equation}\label{ein}
p_\mu p^\mu = \left(\gamma^\mu p_\mu\right)^2 =m^2c^4,
\end{equation}
moves back considerations to the generic Einstein theory with $\epsilon\equiv\infty$, while application of the hyperbolic relation (\ref{ein}) within the modified Dirac equation (\ref{dsi}) leads to two the conventional Dirac theories
\begin{equation}
\left(\gamma^\mu \hat{p}_\mu+M_{\pm}c^2\right)\psi^{\pm}=0,\label{dir}
\end{equation}
where $\psi^\pm$ are the Dirac fields related to the nonhermitian mass matrices $M_{\pm}$, that in general are dependent on an energy $E$ and a mass $m$ of an original quantum relativistic particle. It is of course the case of positive signs in the equation (\ref{dsi}). In fact for any signs case there 2 the Dirac fields, so that the equation (\ref{dsi}) describes 8 the Dirac fields.

With using of the basis of projectors, $M_{\pm}$ can be decomposed as follows
\begin{eqnarray}
  &M_{\pm}=\mu_R^\pm\dfrac{1+\gamma^5}{2}+\mu_L^\pm\dfrac{1-\gamma^5}{2},&\label{mm}\\
  &\mu_R^\pm=-\dfrac{1}{c^2}\left(\dfrac{\epsilon}{2}\pm\sqrt{\strut{\epsilon^2-4\epsilon mc^2-4E^2}}\right),&\label{mu1}\\
  &\mu_L^\pm=\dfrac{1}{c^2}\left(\dfrac{\epsilon}{2}\pm\sqrt{\strut{\epsilon^2+4\epsilon mc^2-4E^2}}\right),&\label{mu2}
\end{eqnarray}
where $\mu^\pm_{R,L}$ are the projected masses, or equivalently can be presented as a sum of its hermitian $\mathfrak{H}(M)$ and antihermitian $\mathfrak{A}(M)$ parts
\begin{eqnarray}
&M_{\pm}=\mathfrak{H}(M_{\pm})+\mathfrak{A}(M_{\pm}),&\label{mm1}\\
&\mathfrak{H}(M_{\pm})=\dfrac{\mu_R^\pm+\mu_L^\pm}{2}\mathbf{1}_4\quad,\quad\mathfrak{A}(M_\pm)=\dfrac{\mu_R^\pm-\mu_L^\pm}{2}\gamma^5.&\label{hah}
\end{eqnarray}
Introducing the chiral right- and left-handed Weyl fields $\psi_{R,L}^\pm$ related to the Dirac field $\psi^\pm$ according to the standard transformation $\psi_{R,L}^\pm=\dfrac{1\pm\gamma^5}{2}\psi^\pm$ one obtains two the massive Weyl equations with diagonal and hermitian mass matrices $\mu_\pm$
\begin{equation}\label{neu}
\left(\gamma^\mu\hat{p}_\mu+\mu_\pm c^2\right)\left[\begin{array}{c}\psi_R^\pm\\\psi_L^\pm\end{array}\right]=0\quad,
\quad\mu^\pm=\left[\begin{array}{cc}\mu_R^\pm&0\\0&\mu_L^\pm\end{array}\right],
\end{equation}
so that we have totally 16 chiral fields describing by the Weyl equations (\ref{neu}) received from the Dirac equations (\ref{dsi}). The massive Weyl theories (\ref{neu}) are the Euler--Lagrange equations of motion for the gauge field theory with chiral symmetry $SU(3)_C^{TOT}=SU(3)_C^+\oplus SU(3)_C^-$
\begin{equation}
  \mathcal{L}=\mathcal{L}^+ + \mathcal{L}^-,
\end{equation}
where $\mathcal{L}^\pm$ are the Lagrangians associated with the fields $\psi^pm_{R,L}$ as follows
\begin{eqnarray}\label{lag}
  \mathcal{L}^\pm=\bar{\psi}^\pm_R\gamma^\mu \hat{p}_\mu\psi_R^\pm + \bar{\psi}_L^\pm\gamma^\mu \hat{p}_\mu\psi_L^\pm+\mu_R^\pm c^2\bar{\psi}_R\psi_R^\pm+\mu_L^\pm c^2\bar{\psi}_L\psi_L^\pm,
\end{eqnarray}
which is spontaneously broken to the composed gauge field theory with the isospin symmetry $SU(2)_V^{TOT}=SU(2)_V^+\oplus SU(2)_V^-$
\begin{eqnarray}
  \mathcal{L}&=&\bar{\psi^+}\left(\gamma^\mu \hat{p}_\mu+\mu_{eff}^+ c^2\right)\psi^++\bar{\psi^-}\left(\gamma^\mu \hat{p}_\mu+\mu_{eff}^-c^2\right)\psi^-=\label{tlag1a}\\
  &=&\bar{\Psi}\left(\gamma^\mu\hat{p}_{\mu}+M_{eff}c^2\right)\Psi,\label{tlag1b}
\end{eqnarray}
where $\mu_{eff}^{\pm}$ are the effective mass matrices of the gauge fields $\psi^\pm$, and $M_{eff}$ is the mass matrix of the effective composed field $\Psi=\left[\begin{array}{c}{\psi^+}\\ {\psi^-}\end{array}\right]$ given by
\begin{eqnarray}
  \mu_{eff}^\pm=\dfrac{\mu_R^\pm-\mu_L^\pm}{2}\gamma^5\quad,\quad M_{eff}=\left[\begin{array}{cc}{\mu^+_{eff}}&0\\0&{\mu^-_{eff}}\end{array}\right].
\end{eqnarray}
The Lagrangian (\ref{tlag1b}) describes the effective field theory -- composed chiral condensate of massive neutrinos.

In this paper we will consider both the Dirac equations (\ref{dir}) and the massive Weyl equations (\ref{neu}), but we are not going to discuss the gauge field theory (\ref{tlag1b}) that will be studying in our next topical papers. We will assume that both the neutrinos masses (\ref{mu1}) and (\ref{mu2}) are real numbers, \emph{i.e.} we will consider situation when the maximal energy $\epsilon$ deforming Special Relativity is determined by a relativistic particle characteristics as follows
\begin{eqnarray}
  \epsilon&\in&\left(-\infty,-2mc^2\left(1+\sqrt{\strut{1+\left(\dfrac{E}{mc^2}\right)^2}}\right)\right]\cup\nonumber\\
  &\cup&\left[-2mc^2\left(1-\sqrt{\strut{1+\left(\dfrac{E}{mc^2}\right)^2}}\right),2mc^2\left(1+\sqrt{\strut{1+\left(\dfrac{E}{mc^2}\right)^2}}\right)\right]\cup\nonumber\\
  &\cup&\left[2mc^2\left(1+\sqrt{\strut{1+\left(\dfrac{E}{mc^2}\right)^2}}\right),\infty\right),
\end{eqnarray}
or for the case of massless relativistic particle possessing an energy $E$ is
\begin{eqnarray}
  \epsilon\in\left(-\infty,-2|E|\right]\cup\left[2|E|,\infty\right).
\end{eqnarray}

\section{Energy renormalization}
In fact existence of the massive neutrinos allows to explain in a consistent physical way the nature of the Snyder noncommutative geometry model. Let us see that by direct elementary algebraic manipulations the relations for masses of left- and right- chiral Weyl fields (\ref{mu1}) and (\ref{mu2}) can be rewritten in the form of the system of equations
\begin{equation}\label{mus}
\left\{\begin{array}{c}
\left(\mu_R^{\pm}c^2+\dfrac{\epsilon}{2}\right)^2=\epsilon^2-4\epsilon mc^2 -4E^2\\
\left(\mu_L^{\pm}c^2-\dfrac{\epsilon}{2}\right)^2=\epsilon^2+4\epsilon mc^2 -4E^2
\end{array}\right.
\end{equation}
which allows to study dependence of the deformation energy parameter $\epsilon$ and the particle energy $E$ from the masses $m,\mu_R^{\pm},\mu_L^{\pm}$ treated as physically measurable quantities. By subtraction of the second equation from the first one (\ref{mus}) one establishes the relation
\begin{equation}
 \left(\mu_L^{\pm}c^2-\dfrac{\epsilon}{2}\right)^2-\left(\mu_R^{\pm}c^2+\dfrac{\epsilon}{2}\right)^2=8\epsilon mc^2,
\end{equation}
which after application of elementary algebraic manipulations allows to derive the deformation energy in the Snyder model (\ref{nd}) as
\begin{equation}\label{def}
  \epsilon=\dfrac{\left(\mu_L^{\pm}-\mu_R^{\pm}\right)c^2}{1-\dfrac{8m}{\mu_L^{\pm}+\mu_R^{\pm}}}.
\end{equation}
A maximal energy (\ref{def}) does not vanish for all $\mu_L^{\pm}\neq\mu_L^{\pm}\neq0$, and is finite for all $\mu_R^{\pm}+\mu_L^{\pm}\neq8m$. Here $m$ is the mass of an original quantum state, and both $\mu_R^{\pm}$ and $\mu_L^{\pm}$ are assumed as physical quantities. In supposition all the masses can be fixed by experiments. In the case, when an original state is massless, one obtains
\begin{equation}
  \epsilon(m=0)=\left(\mu_L^{\pm}-\mu_R^{\pm}\right)c^2\equiv\epsilon_0,
\end{equation}
that is finite and non vanishing for finite $\mu_R^{\pm}\neq0$ and $\mu_L^{\pm}\neq0$. In this manner we have
\begin{equation}
  \epsilon=\epsilon_0\left[1+\dfrac{8m}{\mu_R^{\pm}+\mu_L^{\pm}}+O\left(\left(\dfrac{8m}{\mu_R^{\pm}+\mu_L^{\pm}}\right)^2\right)\right],
\end{equation}
for all $|\mu_R^{\pm}+\mu_L^{\pm}|>8m$, and
\begin{equation}
  \epsilon=\epsilon_0\left[\dfrac{\mu_R^{\pm}+\mu_L^{\pm}}{8m}+O\left(\left(\dfrac{\mu_R^{\pm}+\mu_L^{\pm}}{8m}\right)^2\right)\right],
\end{equation}
for all $|\mu_R^{\pm}+\mu_L^{\pm}|<8m$. On the other hand, however, addition of the second equation to the first one in (\ref{mus}) gives the relation
\begin{equation}\label{c2}
 \left(\mu_L^{\pm}c^2-\dfrac{\epsilon}{2}\right)^2+\left(\mu_R^{\pm}c^2+\dfrac{\epsilon}{2}\right)^2=2\left(\epsilon^2-4E^2\right),
\end{equation}
which can be treated as the constraint for the energy $E$ of a relativistic particle, immediately solved with respect $E$, and presented in the canonical quadratic form with respect to the energy parameter $\epsilon$
\begin{equation}\label{er}
 E^2=\dfrac{3}{16}\left\{\left[\epsilon+\dfrac{\mu_L^{\pm}-\mu_R^{\pm}}{3}c^2\right]^2-\left[\dfrac{\mu_L^{\pm}-\mu_R^{\pm}}{3}c^2\right]^2\left[7+\dfrac{12\mu_L^{\pm}\mu_R^{\pm}}{\left(\mu_L^{\pm}-\mu_R^{\pm}\right)^2}\right]\right\}.
\end{equation}
By application of the deformation parameter energy (\ref{def}) into the energetic constraint of a relativistic particle (\ref{er}) one obtains the formula
\begin{equation}\label{er1}
 E^2=\dfrac{\left[\left(\mu_L^{\pm}-\mu_R^{\pm}\right)c^2\right]^2}{48}\left\{\left(\dfrac{4-\dfrac{8m}{\mu_L^{\pm}+\mu_R^{\pm}}}{1-\dfrac{8m}{\mu_L^{\pm}+\mu_R^{\pm}}}\right)^2-\left[7+\dfrac{12\mu_L^{\pm}\mu_R^{\pm}}{\left(\mu_L^{\pm}-\mu_R^{\pm}\right)^2}\right]\right\},
\end{equation}
which for the case of originally massless state reduces into the form
\begin{equation}\label{e0}
  E^2(m=0)=\dfrac{1}{16}\left[\left(\mu_L^{\pm}-\mu_R^{\pm}\right)c^2\right]^{2}\left[3-4\dfrac{\mu_L^{\pm}\mu_R^{\pm}}{\left(\mu_L^{\pm}-\mu_R^{\pm}\right)^2}\right]\equiv E^2_0.
\end{equation}
In fact, for given $E_0$ the equation (\ref{e0}) can be used for establishment of the relation between masses of the neutrinos. In result one receives two possible solutions
\begin{equation}
  \mu_R^{\pm}=\dfrac{4}{3}\mu_L^{\pm}\left[\dfrac{5}{4}\pm\sqrt{\strut{1+3\left(\dfrac{\mu_0}{\mu_L^{\pm}}\right)^2}}\right]\quad,\quad \mu_0\equiv \dfrac{E_0}{c^2},
\end{equation}
which are minimized for the value $\mu_0\equiv0$ by the values
\begin{equation}
  \mu_R^{\pm}=\left\{3\mu_L^{\pm},\dfrac{1}{3}\mu_L^{\pm}\right\}.
\end{equation}
Interestingly, there is a possibility of the one solution between the masses $\mu_L^{\pm}$ and $\mu_R^{\pm}$ that is given by putting $\mu_0$ as a tachyonic mass
\begin{equation}
  \mu_0=i\dfrac{\mu_L^{\pm}}{\sqrt{3}},
\end{equation}
and results in the relation
\begin{equation}
  \mu_R^{\pm}=\dfrac{5}{3}\mu_L^{\pm}.
\end{equation}

For all $|\mu_R^{\pm}+\mu_L^{\pm}|>8m$ the constraint (\ref{er1}) can be approximated by
\begin{equation}\label{er2}
 E^2-E_0^2=\left(\dfrac{\mu_L^{\pm}-\mu_R^{\pm}}{2}c^2\right)^2\left[2\dfrac{8m}{\mu_L^{\pm}+\mu_R^{\pm}}+O\left[\left(\dfrac{8m}{\mu_L^{\pm}+\mu_R^{\pm}}\right)^2\right]\right],
\end{equation}
and for $|\mu_R^{\pm}+\mu_L^{\pm}|<8m$ the leading approximation is
\begin{equation}\label{er3}
 E^2-E_0^2=\left(\dfrac{\mu_L^{\pm}-\mu_R^{\pm}}{2}c^2\right)^2\left[-\dfrac{5}{4}-\dfrac{1}{2}\dfrac{8m}{\mu_L^{\pm}+\mu_R^{\pm}}+O\left[\left(\dfrac{8m}{\mu_L^{\pm}+\mu_R^{\pm}}\right)^2\right]\right],
\end{equation}
From the relation (\ref{c2}) one sees that, because the LHS as a sum of two squares of real numbers is always positive, it follows that the RHS must be positive also. In result we obtain the renormalization of a relativistic particle's energy $E$ values
\begin{equation}\label{reg}
  -\dfrac{\epsilon}{2}\leqslant E\leqslant\dfrac{\epsilon}{2},
\end{equation}
Naturally, for the generic case of Special Relativity we have $\epsilon\equiv\infty$ and by this energy $E$ values are not limited. In this manner, in fact the Snyder noncommutative geometry results in energy renormalization of a relativistic particle.
\section{Integrability I: The Dirac equation}
The modified Dirac equation (\ref{dir}) can be straightforwardly rewritten in the Schr\"odinger evolutionary equation form (See \emph{e.g.} the papers \cite{schrod} and the books \cite{quantum})
\begin{equation}\label{scd}
  i\hbar\partial_0\psi^\pm=\hat{H}\psi^\pm,
\end{equation}
where in the present case the Hamilton operator $\hat{H}$ can be established as
\begin{equation}\label{ham}
  \hat{H}=-i\hbar c\gamma^0\gamma^i\partial_i-\dfrac{\mu_L^{\pm}+\mu_R^{\pm}}{2}c^2\gamma^0+\dfrac{\mu_L^{\pm}-\mu_R^{\pm}}{2}c^2\gamma^0\gamma^5,
\end{equation}
and splitted into its hermitian $\mathfrak{H}(\hat{H})$ and antihermitian $\mathfrak{A}(\hat{H})$ parts
\begin{eqnarray}
\hat{H}&=&\mathfrak{H}(\hat{H})+\mathfrak{A}(\hat{H}),\label{hamm}\\
\mathfrak{H}(\hat{H})&=&-i\hbar c\gamma^0\gamma^i\partial_i-\dfrac{\mu_L^{\pm}+\mu_R^{\pm}}{2}c^2\gamma^0,\label{hp}\\
\mathfrak{A}(\hat{H})&=&\dfrac{\mu_L^{\pm}-\mu_R^{\pm}}{2}c^2\gamma^0\gamma^5,\label{ahp}
\end{eqnarray}
with (anti)hermiticity defined standardly
\begin{eqnarray}
  \int d^3x \bar{\psi^\pm}\mathfrak{H}(\hat{H})\psi^\pm &=&\int d^3x \overline{\mathfrak{H}(\hat{H})\psi^\pm}\psi^\pm,\\
  \int d^3x \bar{\psi^\pm}\mathfrak{A}(\hat{H})\psi^\pm &=&-\int d^3x \overline{\mathfrak{A}(\hat{H})\psi^\pm}\psi^\pm.
\end{eqnarray}
Note that in the case of equal masses $\mu_R^{\pm}=\mu_L^{\pm}\equiv\mu$ the antihermitian part (\ref{ahp}) vanishes identically, so that the hermitian one (\ref{hp}) gives the full contribution, and consequently (\ref{hamm}) becomes the usual Dirac Hamiltonian
\begin{equation}
  \hat{H}_{D}=-\gamma^0\left(i\hbar c\gamma^i\partial_i+\mu c^2\right).\label{diru}
\end{equation}
For this usual case, however, from (\ref{def}) one concludes that
\begin{equation}
  \epsilon\equiv 0,
\end{equation}
so in fact the minimal scale becomes infinite formally $\ell\equiv \infty$, and by (\ref{er1}) relativistic particle's energy becomes $E=i\dfrac{1}{2}\mu c^2$ with some mass $\mu$. If we, however, take into account the tachyonic mass case $\mu\rightarrow i\mu=\mu'$ then (\ref{diru}) becomes
\begin{equation}
  \hat{H}_{D}=-\gamma^0\left(i\hbar c\gamma^i\partial_i+i\mu' c^2\right),\label{diru1}
\end{equation}
and $E\equiv\dfrac{1}{2}\mu'c^2$. The relation (\ref{reg}), however, is not validate in this case.

The full modified Hamiltonian (\ref{ham}) has nonhermitian character evidently, so consequently the time evolution (\ref{scd}) is manifestly non unitary. Its formal integration, however, can be carried out in the usual way with the following time evolution operator
\begin{equation}
  \psi^\pm(x,t)=G(t,t_0)\psi^\pm(x,t_0)\quad,\quad G(t,t_0)\equiv\exp\left\{-\dfrac{i}{\hbar}\int_{t_0}^td\tau\hat{H}(\tau)\right\}.\label{int}
\end{equation}
By this reason, the integrability problem for (\ref{scd}) is contained in the appropriate Zassenhaus formula
\begin{eqnarray}
\exp\left\{A+B\right\}&=&\exp(A)\exp(B)\prod_{n=2}^\infty \exp{C_n},\label{zassen}\\
C_2&=&-\frac{1}{2}C,\\
C_3&=&-\frac{1}{6}(2[C,B]+[C,A]),\\
C_4&=&-\frac{1}{24}([[C,A],A]+3[[C,A],B]+3[[C,B],B]),\\
&\ldots&\nonumber
\end{eqnarray}
where $C=[A,B]$. By the formula (\ref{int}) one has identification
\begin{eqnarray}
  A&\equiv&A(t)=-\frac{i}{\hbar}\int_{t_0}^t d\tau\mathfrak{H}(\hat{H})(\tau),\\
  B&\equiv&B(t)=-\frac{i}{\hbar}\int_{t_0}^t d\tau \mathfrak{A}(\hat{H})(\tau),
\end{eqnarray}
so that the commutator $C$ is established as
\begin{equation}
  C=-\dfrac{1}{\hbar^2}\int_{t_0}^t d\tau' \int_{t_0}^t d\tau''\mathfrak{C}\left(\tau',\tau''\right),
\end{equation}
where
\begin{equation}
  \mathfrak{C}\left(\tau',\tau''\right)\equiv\left[\mathfrak{H}(\hat{H})(\tau'),\mathfrak{A}(\hat{H})(\tau'')\right].
\end{equation}
Straightforward calculation of $\mathfrak{C}$ can be done by elementary algebra
\begin{eqnarray}
\mathfrak{C}&=&\left(i\hbar\dfrac{\mu_R^{\pm}-\mu_L^{\pm}}{2}c^3\partial_i\right)\gamma^0\gamma^i\gamma^0\gamma^5+\left(\dfrac{(\mu_R^{\pm})^2-(\mu_L^{\pm})^2}{4}c^4\right)\gamma^0\gamma^0\gamma^5-\\
&-&\left(i\hbar\dfrac{\mu_R^{\pm}-\mu_L^{\pm}}{2}c^3\partial_i\right)\gamma^0\gamma^5\gamma^0\gamma^i-\left(\dfrac{(\mu_R^{\pm})^2-(\mu_L^{\pm})^2}{4}c^4\right)\gamma^0\gamma^5\gamma^0=\\
&=&2\left(i\hbar\dfrac{\mu_R^{\pm}-\mu_L^{\pm}}{2}c^3\partial_i\right)\gamma^0\gamma^i\gamma^0\gamma^5+2\left(\dfrac{(\mu_R^{\pm})^2-(\mu_L^{\pm})^2}{4}c^4\right)\gamma^0\gamma^0\gamma^5,
\end{eqnarray}
where we have applied the relations
\begin{equation}
\gamma^0\gamma^5\gamma^0\gamma^i=-\gamma^0\gamma^i\gamma^0\gamma^5\quad,\quad \gamma^0\gamma^5\gamma^0=-\gamma^0\gamma^0\gamma^5,
\end{equation}
arising by employing the usual Clifford algebra of the Dirac matrices
\begin{equation}
\left\{\gamma^\mu, \gamma^\nu\right\}=2\eta^{\mu\nu}\mathbf{1}_4\quad,\quad\left\{\gamma^5, \gamma^\mu\right\}=0\quad,\quad\gamma^5=i\gamma^0\gamma^1\gamma^2\gamma^3.
\end{equation}
So, consequently one obtains the result
\begin{equation}
  \mathfrak{C}(\tau',\tau'')=2\mathfrak{H}(\hat{H})(\tau')\mathfrak{A}(\hat{H})(\tau''),
\end{equation}
that leads to the equivalent statement -- for any times $\tau'$ and $\tau''$ the Poisson bracket between the hermitian $\mathfrak{H}(\hat{H})(\tau')$ and the antihermitian $\mathfrak{A}(\hat{H})(\tau'')$ parts of the total Hamiltonian (\ref{hamm}) is trivial
\begin{equation}
  \left\{\mathfrak{H}(\hat{H})(\tau'),\mathfrak{A}(\hat{H})(\tau'')\right\}=0.
\end{equation}
Naturally, by simple factorization one obtains also
\begin{equation}
  C=2AB\quad,\quad\{A,B\}=0,
\end{equation}
and consequently
\begin{eqnarray}
  \left[C,A\right]&=&CA,\\
  \left[C,B\right]&=&CB,\\
  \left[\left[C,A\right],A\right]&=&2\left[C,A\right]A,\\
  \left[\left[C,A\right],B\right]&=&2\left[C,A\right]B,\\
  \left[\left[C,B\right],A\right]&=&2\left[C,B\right]A,
\end{eqnarray}
and so on. In result the 4th order approximation of the Zassehnaus formula (\ref{zassen}) in the present case is
\begin{eqnarray}
\exp\left\{A+B\right\}&\approx&\exp(A)\exp(B)\exp{C_2}\exp{C_3}\exp{C_4},\\
C_2&=&-\frac{1}{2}C,\\
C_3&=&-\frac{1}{6}(CA+2CB),\\
C_4&=&-\frac{1}{12}\left(CA^2+3CB^2+\dfrac{3}{2}C^2\right).
\end{eqnarray}
For the case of constant in time masses $\mu_R^{\pm}$ and $\mu_L^{\pm}$ one determine the relations
\begin{eqnarray}
  A&=&\dfrac{i}{\hbar}(t-t_0)\left(-i\hbar c \gamma^i\partial_i+\dfrac{\mu_L^{\pm}+\mu_R^{\pm}}{2}c^2\right)\gamma^0,\\
  B&=&\dfrac{i(\mu_L^{\pm}-\mu_R^{\pm})c^2}{2\hbar}(t-t_0)\gamma^5\gamma^0,\\
  C&=&\dfrac{(\mu_L^{\pm}-\mu_R^{\pm})c^2}{\hbar^2}(t-t_0)^2\left(-i\hbar c\gamma^i\partial_i+\dfrac{\mu_L^{\pm}+\mu_R^{\pm}}{2}c^2\right)\gamma^5,
\end{eqnarray}
and consequently by elementary algebraic manipulations one establishes the Zassenhaus exponents as
\begin{eqnarray}
C_2&=&-\dfrac{(\mu_L^{\pm}-\mu_R^{\pm})c^2}{2\hbar^2}(t-t_0)^2\left(-i\hbar c\gamma^i\partial_i+\dfrac{\mu_L^{\pm}+\mu_R^{\pm}}{2}c^2\right)\gamma^5,\\
C_3&=&-\dfrac{i}{6\hbar^3}(\mu_L^{\pm}-\mu_R^{\pm})c^2(t-t_0)^3\left(-i\hbar c\gamma^i\partial_i+\dfrac{\mu_L^{\pm}+\mu_R^{\pm}}{2}c^2\right)\times\nonumber\\
&\times&\left[\left(-i\hbar c\gamma^i\partial_i+\dfrac{\mu_L^{\pm}+\mu_R^{\pm}}{2}c^2\right)\gamma^5+(\mu_L^{\pm}-\mu_R^{\pm})c^2\right]\gamma^0,\\
C_4&=&\dfrac{(\mu_L^{\pm}-\mu_R^{\pm})c^2}{12\hbar^4}(t-t_0)^4\left(-i\hbar c\gamma^i\partial_i+\dfrac{\mu_L^{\pm}+\mu_R^{\pm}}{2}c^2\right)\times\nonumber\\
&\times&\Bigg\{\left[\left(-i\hbar c\gamma^i\partial_i+\dfrac{\mu_L^{\pm}+\mu_R^{\pm}}{2}c^2\right)^2+3\left(\dfrac{\mu_L^{\pm}-\mu_R^{\pm}}{2}c^2\right)^2\right]\gamma^5+\nonumber\\
&+&3\dfrac{\mu_L^{\pm}-\mu_R^{\pm}}{2}c^2\left(-i\hbar c\gamma^i\partial_i+\dfrac{\mu_L^{\pm}+\mu_R^{\pm}}{2}c^2\right)\Bigg\}.
\end{eqnarray}
Exponents $C_n$ show in a manifest way that the integrability problem is not well defined. Namely, the Zassenhaus coefficients $C_n$ are still a sums of two noncommuting operators. The fundamental stage, \emph{i.e.} the exponentialization procedure, must be applied again, so that consequently in the next step one has the same problem, \emph{i.e.} the cyclic problem. Therefore this recurrence is not algorithm, that is the symptom of non integrability of (\ref{scd}).
\section{Integrability II: The Weyl equation}
For solving the problem, let us consider the integrability procedure with respect to the massive Weyl equation (\ref{neu}). This equation can be straightforwardly rewritten in the form of the Schr\"odinger time evolution
\begin{equation}\label{qd}
  i\hbar\partial_0\left[\begin{array}{c}\psi^\pm_R(x,t)\\\psi^\pm_L(x,t)\end{array}\right]=\hat{H}\left(\partial_i\right)\left[\begin{array}{c}\psi^\pm_R(x,t)\\\psi^\pm_L(x,t)\end{array}\right],
\end{equation}
where the hermitian Hamilton operator $\hat{H}$ defines to the unitary evolution
\begin{equation}\label{ham2}
  \hat{H}=-\gamma^0\left(i\hbar c\gamma^i\partial_i+\left[\begin{array}{cc}\mu_R^{\pm}c^2&0\\0&\mu_L^{\pm}c^2\end{array}\right]\right),
\end{equation}
so that the integration can be done in the usual quantum mechanical way. Integrability of (\ref{qd}) is well defined. The solutions are
\begin{equation}
  \left[\begin{array}{c}\psi^\pm_R(x,t)\\\psi^\pm_L(x,t)\end{array}\right]=U(t,t_0)\left[\begin{array}{c}\psi^\pm_R(x,t_0)\\\psi^\pm_L(x,t_0)\end{array}\right],
\end{equation}
where $U(t,t_0)$ is the unitary time-evolution operator, that for the constant masses is explicitly given by
\begin{equation}\label{uexp}
  U(t,t_0)=\exp\left\{-\dfrac{i}{\hbar}(t-t_0)\hat{H}\right\},
\end{equation}
and $\psi^\pm_{R,L}(x,t_0)$ are the initial time $t_0$ eigenstates with defined momenta
\begin{equation}\label{peig}
  i\hbar \sigma^i\partial_i\psi^\pm_{R,L}(x,t_0)={p_{R,L}^\pm}^0\psi^\pm_{R,L}(x,t_0),
\end{equation}
where the momenta ${p_{R}^\pm}^0$ and ${p_{L}^\pm}^0$ are related to the right- $\psi^\pm_{R}(x,t_0)$ or left-handed $\psi^\pm_{L}(x,t_0)$ chiral fields, respectively. The eigenequation (\ref{peig}), however, can be straightforwardly integrated. The result can be presented in the compact form
\begin{equation}\label{psix}
\psi^\pm_{R,L}(x,t_0)=\exp\left\{-\dfrac{i}{\hbar}{p_{R,L}^\pm}^0(x-x_0)_i\sigma^i\right\}\psi^\pm_{R,L}(x_0,t_0),
\end{equation}
or after direct exponentialization
\begin{eqnarray}\label{psix1}
  &&\psi^\pm_{R,L}(x,t_0)=\Bigg\{\mathbf{1}_2\cos\left|\dfrac{{p_{R,L}^\pm}^0}{\hbar}(x-x_0)_i\right|-\nonumber\\
  &-&i\left[\dfrac{{p_{R,L}^\pm}^0}{\hbar}(x-x_0)_i\sigma^i\right]\dfrac{\sin\left|\dfrac{{p_{R,L}^\pm}^0}{\hbar}(x-x_0)_i\right|}{\left|\dfrac{{p_{R,L}^\pm}^0}{\hbar}(x-x_0)_i\right|}\Bigg\}\psi^\pm_{R,L}(x_0,t_0).
\end{eqnarray}
 Currently the embarrassing problem presented in the integration procedure of the Dirac equation, discussed in the previous section, is absent. The Zessenhaus formula is not troublesome now because the matrix $\gamma^5$ is by definition included into the Weyl fields, so that the Hamilton operator (\ref{ham2}) is pure hermitian, and consequently the exponentialization (\ref{uexp}) can be done is the usual way. At first glance, however, the mass matrix presence in the Hamilton operator (\ref{ham2}) causes that one chooses at least two nonequivalent representations of the Dirac $\gamma$ matrices. Straightforward analogy to the massless Weyl equation says that the appropriate choice is the Weyl basis. On the other hand, however, the Hamilton operator (\ref{ham2}) is usual hermitian Dirac Hamiltonian, so consequently the Dirac basis is the right representation. In this manner, in fact, we should consider rather both the chiral fields and the time evolution operator (\ref{uexp}) strictly related the chosen representation $(r)$
 \begin{eqnarray}
   U(t,t_0)&\rightarrow&U^{r}(t,t_0),\\
   \psi^\pm_{R,L}(x,t_0)&\rightarrow&(\psi^\pm_{R,L})^{r}(x,t_0),\\
   \psi^\pm_{R,L}(x_0,t_0)&\rightarrow&(\psi^\pm_{R,L})^{r}(x_0,t_0)
 \end{eqnarray}
 where the upper index $r=D,W$ means that the quantities are taken in the Dirac or the Weyl basis. The eigenequation (\ref{peig}), however, is independent on the representation choice, so that it physical condition - the fields have measurable momenta ${p_{R,L}^\pm}^0$. For full correctness, let us test both choices.
\subsection{The Dirac basis}
The Dirac basis of the gamma matrices is defined as
\begin{equation}\label{dirrep}
  \gamma^0=\left[\begin{array}{cc}I&0\\0&-I\end{array}\right]\quad,\quad\gamma^i=\left[\begin{array}{cc}0&\sigma^i\\-\sigma^i&0\end{array}\right]\quad,\quad\gamma^5=\left[\begin{array}{cc}0&I\\I&0\end{array}\right],
\end{equation}
where $I$ is the $2\times2$ unit matrix, and $\sigma^i=[\sigma_x,\sigma_y,\sigma_z]$ is a vector of the $2\times2$ Pauli matrices
\begin{equation}\label{pauli}
\sigma_x=\left[\begin{array}{cc}0&1\\1&0\end{array}\right]\quad,\quad\sigma_y=\left[\begin{array}{cc}0&-i\\i&0\end{array}\right]\quad,\quad\sigma_z=\left[\begin{array}{cc}1&0\\0&-1\end{array}\right].
\end{equation}
Consequently, by using of (\ref{dirrep}) the Hamilton operator (\ref{ham2}) becomes
\begin{equation}\label{ham2a}
  \hat{H}=\left[\begin{array}{cc}\mu_R^{\pm}&i\dfrac{\hbar}{c}\sigma^i\partial_i\\
  i\dfrac{\hbar}{c}\sigma^i\partial_i&-\mu_L^{\pm}\end{array}\right]c^2,
\end{equation}
and for the case of constant in time neutrinos masses yields a solution by the unitary time evolution operator $U$
\begin{equation}
  U^D=\exp\left\{-i\dfrac{c^2}{\hbar}(t-t_0)\left[\begin{array}{cc}\mu_R^{\pm}&i\dfrac{\hbar}{c}\sigma^i\partial_i\\
  i\dfrac{\hbar}{c}\sigma^i\partial_i&-\mu_L^{\pm}\end{array}\right]\right\}.\label{U}
\end{equation}
After straightforward exponentialization (\ref{U}) can be written in the compact form
\begin{eqnarray}\label{dteo}
U^D&=&\Bigg\{\left[\begin{array}{cc}I&0\\0&I\end{array}\right]
  \cos\left[\dfrac{t-t_0}{\hbar}c^2\sqrt{\strut{\left(\dfrac{\mu_R^{\pm}+\mu_L^{\pm}}{2}\right)^2}+\left(i\dfrac{\hbar}{c}\sigma^i\partial_i\right)^2}\right]-\nonumber\\
  &-&i\left[\begin{array}{cc}\dfrac{\mu_L^{\pm}+\mu_R^{\pm}}{2}&i\dfrac{\hbar}{c}\sigma^i\partial_i\\i\dfrac{\hbar}{c}\sigma^i\partial_i&-\dfrac{\mu_L^{\pm}+\mu_R^{\pm}}{2}\end{array}\right]\times\nonumber\\
  &\times&\dfrac{\sin\left[\dfrac{t-t_0}{\hbar}c^2\sqrt{\strut{\left(\dfrac{\mu_R^{\pm}+\mu_L^{\pm}}{2}\right)^2+\left(i\dfrac{\hbar}{c}\sigma^i\partial_i\right)^2}}\right]}{\sqrt{\strut{\left(\dfrac{\mu_R^{\pm}+\mu_L^{\pm}}{2}\right)^2+\left(i\dfrac{\hbar}{c}\sigma^i\partial_i\right)^2}}}\Bigg\}\times\nonumber\\
  &\times&\exp\left\{-i\dfrac{(\mu_R^{\pm}-\mu_L^{\pm})c^2}{2\hbar}(t-t_0)\right\},
\end{eqnarray}
where we understand the all the functions are treated by the appropriate Taylor series expansions.
\subsection{The Weyl basis}
Equivalently, however, one can consider employing of the Weyl representation of the Dirac $\gamma$ matrices. This basis is defined as
\begin{equation}\label{wrep}
  \gamma^0=\left[\begin{array}{cc}0&I\\I&0\end{array}\right]\quad,\quad\gamma^i=\left[\begin{array}{cc}0&\sigma^i\\-\sigma^i&0\end{array}\right]\quad,\quad\gamma^5=\left[\begin{array}{cc}-I&0\\0&I\end{array}\right].
\end{equation}
For the choice of a representation in the form (\ref{wrep}) the massive Weyl equation (\ref{qd}) is governed by the Hamilton operator (\ref{ham2}) having the following form
\begin{equation}\label{ham2b}
  \hat{H}=\left[\begin{array}{cc}i\dfrac{\hbar}{c}\sigma^i\partial_i&-\mu_L^{\pm}\\
  -\mu_R^{\pm}&-i\dfrac{\hbar}{c}\sigma^i\partial_i\end{array}\right]c^2.
\end{equation}
Consequently, for the case of constant in time neutrinos masses one establishes the unitary time evolution operator in the following formal form
\begin{equation}
  U^W=\exp\left\{-i\dfrac{c^2}{\hbar}(t-t_0)\left[\begin{array}{cc}i\dfrac{\hbar}{c}\sigma^i\partial_i&-\mu_L^{\pm}\\
  -\mu_R^{\pm}&-i\dfrac{\hbar}{c}\sigma^i\partial_i\end{array}\right]\right\},\label{U1}
\end{equation}
which after straightforward elementary exponentialization procedure can be presented in the form
\begin{eqnarray}\label{wteo}
  U^W&=&\left[\begin{array}{cc}I&0\\0&I\end{array}\right]
  \cos\left[\dfrac{t-t_0}{\hbar}c^2\sqrt{\strut{\mu_L^{\pm}\mu_R^{\pm}+\left(i\dfrac{\hbar}{c}\sigma^i\partial_i\right)^2}}\right]-\nonumber\\
  &-&i\left[\begin{array}{cc}i\dfrac{\hbar}{c}\sigma^i\partial_i&-\mu_L^{\pm}\\-\mu_R^{\pm}&-i\dfrac{\hbar}{c}\sigma^i\partial_i\end{array}\right]\times\nonumber\\
  &\times&\dfrac{\sin\left[\dfrac{t-t_0}{\hbar}c^2\sqrt{\strut{\mu_L^{\pm}\mu_R^{\pm}+\left(i\dfrac{\hbar}{c}\sigma^i\partial_i\right)^2}}\right]}{\sqrt{\strut{\mu_L^{\pm}\mu_R^{\pm}+\left(i\dfrac{\hbar}{c}\sigma^i\partial_i\right)^2}}}.
\end{eqnarray}

Evidently, time evolution operator derived in the Weyl representation (\ref{wteo}) has simpler form then its Dirac's equivalent (\ref{dteo}). In this way the choices are not physically equivalent, \emph{i.e.} will yield different solutions of the same equation. It is not, however, the strangest result. Namely, both the choices can be related to physics in different energy regions. So that it is useful to solve the massive Weyl equation in both mentioned representations. It must be emphasized that strictly speaking the results obtained in this subsection are related to the massive Weyl equations presented in the Schr\"odinger time-evolution form (\ref{qd}).
\subsection{The space-time evolution}
Presently, one can employ the results received above, \emph{i.e.} the momentum eigenequations (\ref{peig}), the spatial evolutions (\ref{psix}), and the unitary time evolution operators (\ref{dteo}) and (\ref{wteo}), for an exact determination of the appropriate solutions of the massive Weyl equation (\ref{qd}) in both the Dirac and the Weyl representations of the Dirac gamma matrices.
\subsubsection{Dirac-like solutions}
Applying first the Dirac representation, by elementary algebraic manipulations one receives straightforwardly the right-handed chiral Weyl fields in the following form
\begin{eqnarray}\label{dir1s}
&&(\psi^\pm_R)^D(x,t)=\Bigg\{\Bigg[\cos\left[\dfrac{t-t_0}{\hbar}E^D({p_R^\pm}^0)\right]-\nonumber\\
&-&i\mu_{\pm}^Dc^2\dfrac{\sin\left[\dfrac{t-t_0}{\hbar}E^D({p_R^\pm}^0)\right]}{E^D({p_R^\pm}^0)}\Bigg]\exp\left\{-\dfrac{i}{\hbar}{p_{R}^\pm}^0(x-x_0)_i\sigma^i\right\}(\psi^\pm_{R})^{D}_0-\nonumber\\
&-&i{p_{L}^\pm}^0c\dfrac{\sin\left[\dfrac{t-t_0}{\hbar}E^D({p_L^\pm}^0)\right]}{E^D({p_L^\pm}^0)}\exp\left\{-\dfrac{i}{\hbar}{p_{L}^\pm}^0(x-x_0)_i\sigma^i\right\}(\psi^\pm_{L})^{D}_0\Bigg\}\times\nonumber\\
&\times&\exp\left\{-i\dfrac{(\mu_R^{\pm}-\mu_L^{\pm})c^2}{2\hbar}(t-t_0)\right\},
\end{eqnarray}
where for shorten notation $(\psi^\pm_{R,L})^{D}_0=(\psi^\pm_{R,L})^{D}(x_0,t_0)$, ${\mu_\pm^D}=\dfrac{\mu_R^{\pm}+\mu_L^{\pm}}{2}$ and
\begin{equation}
E^D({p_R^\pm}^0)\equiv c^2\sqrt{\strut{\left(\mu_\pm^D\right)^2+\left(\dfrac{{p_{R}^\pm}^0}{c}\right)^2}}.
\end{equation}
Similarly, the left-handed chiral Weyl fields also can be determined in an exact way, the result is analogical to the right-handed case
\begin{eqnarray}\label{dir2s}
&&(\psi^\pm_L)^D(x,t)=\Bigg\{\Bigg[\cos\left[\dfrac{t-t_0}{\hbar}E^D({p_L^\pm}^0)\right]+\nonumber\\
&+&i\mu_{\pm}^Dc^2\dfrac{\sin\left[\dfrac{t-t_0}{\hbar}E^D({p_L^\pm}^0)\right]}{E^D({p_L^\pm}^0)}\Bigg]\exp\left\{-\dfrac{i}{\hbar}{p_{L}^\pm}^0(x-x_0)_i\sigma^i\right\}(\psi^\pm_{L})^{D}_0-\nonumber\\
&-&i{p_{R}^\pm}^0c\dfrac{\sin\left[\dfrac{t-t_0}{\hbar}E^D({p_R^\pm}^0)\right]}{E^D({p_R^\pm}^0)}\exp\left\{-\dfrac{i}{\hbar}{p_{R}^\pm}^0(x-x_0)_i\sigma^i\right\}(\psi^\pm_{R})^{D}_0\Bigg\}\times\nonumber\\
&\times&\exp\left\{-i\dfrac{(\mu_R^{\pm}-\mu_L^{\pm})c^2}{2\hbar}(t-t_0)\right\}.
\end{eqnarray}
\subsubsection{Weyl-like solutions}
Similar line of thought can be carried out in the Weyl basis. An elementary calculation leads to the right-hand chiral Weyl fields in the form
\begin{eqnarray}\label{weyl1s}
&&(\psi^\pm_R)^W(x,t)=\Bigg\{\cos\left[\dfrac{t-t_0}{\hbar}E^W({p_R^\pm}^0)\right]-\nonumber\\
&-&i{p^\pm_R}^0c\dfrac{\sin\left[\dfrac{t-t_0}{\hbar}E^W({p_R^\pm}^0)\right]}{E^W({p_R^\pm}^0)}\Bigg\}\exp\left\{-\dfrac{i}{\hbar}{p_{R}^\pm}^0(x-x_0)_i\sigma^i\right\}(\psi^\pm_{R})^{W}_0+\nonumber\\
&+&i\mu_L^{\pm}c^2\dfrac{\sin\left[\dfrac{t-t_0}{\hbar}E^W({p_L^\pm}^0)\right]}{E^W({p_L^\pm}^0)}\exp\left\{-\dfrac{i}{\hbar}{p_{L}^\pm}^0(x-x_0)_i\sigma^i\right\}(\psi^\pm_{L})^{W}_0,
\end{eqnarray}
where similarly as in the Dirac-like case we have introduced the shorten notation $(\psi^\pm_{R,L})^{W}_0=(\psi^\pm_{R,L})^{W}(x_0,t_0)$, ${\mu_\pm^W}=\mu_{\pm}^W=\sqrt{\strut{\mu_L^{\pm}\mu_R^{\pm}}}$ and
\begin{equation}
  E^W({p_R^\pm}^0)\equiv c^2\sqrt{\strut{\left(\mu_{\pm}^W\right)^2+\left(\dfrac{{p_R^\pm}^0}{c}\right)^2}}.
\end{equation}
For the left-hand chiral Weyl fields one obtains the formula
\begin{eqnarray}\label{weyl2s}
&&(\psi^\pm_L)^W(x,t)=\Bigg\{\cos\left[\dfrac{t-t_0}{\hbar}E^W({p_L^\pm}^0)\right]-\nonumber\\
&+&i{p^\pm_L}^0c\dfrac{\sin\left[\dfrac{t-t_0}{\hbar}E^W({p_L^\pm}^0)\right]}{E^W({p_L^\pm}^0)}\Bigg\}\exp\left\{-\dfrac{i}{\hbar}{p_{L}^\pm}^0(x-x_0)_i\sigma^i\right\}(\psi^\pm_{L})^{W}_0+\nonumber\\
&+&i\mu_R^{\pm}c^2\dfrac{\sin\left[\dfrac{t-t_0}{\hbar}E^W({p_R^\pm}^0)\right]}{E^W({p_R^\pm}^0)}\exp\left\{-\dfrac{i}{\hbar}{p_{R}^\pm}^0(x-x_0)_i\sigma^i\right\}(\psi^\pm_{R})^{W}_0.
\end{eqnarray}

In this manner one sees that the difference between obtained solutions is crucial. Direct comparing of the Weyl-like solutions (\ref{weyl1s}) and (\ref{weyl2s}) with the Dirac-like solutions (\ref{dir1s}) and (\ref{dir2s}) shows that in the Dirac basis case there are different coefficients of cosinuses and sinuses, there is additional time-exponent, and moreover the functions $M^D({p_R^\pm}^0)$ and $M^W({p_R^\pm}^0)$ having a basic status for both the solutions also have different form with respect to choice of the Dirac $\gamma$ matrices representation.
\subsection{Probability density. Normalization}
If we know the chiral Weyl fields, then in the Dirac representation, one can derive the usual Dirac fields by the following way
\begin{equation}\label{pdirac}
  (\psi^\pm)^D=\left[\begin{array}{cc}\dfrac{(\psi^\pm_R)^D+(\psi^\pm_L)^D}{2}\mathbf{1}_2&\dfrac{(\psi^\pm_R)^D-(\psi^\pm_L)^D}{2}\mathbf{1}_2\\
  \dfrac{(\psi^\pm_R)^D-(\psi^\pm_L)^D}{2}\mathbf{1}_2&\dfrac{(\psi^\pm_R)^D+(\psi^\pm_L)^D}{2}\mathbf{1}_2\end{array}\right],
\end{equation}
where for shorten notation $(\psi^\pm)^D=(\psi^\pm)^D(x,t)$, and $(\psi^\pm_{R,L})^D=(\psi^\pm_{R,L})^D(x,t)$. Similarly in the Weyl representation, the Dirac fields can be determined as
\begin{equation}\label{pweyl}
  (\psi^\pm)^W=\left[\begin{array}{cc}(\psi^\pm_L)^W\mathbf{1}_2&\mathbf{0}_2\\
  \mathbf{0}_2&(\psi^\pm_R)^W\mathbf{1}_2\end{array}\right],
\end{equation}
where also we have used the shorten notation $(\psi^\pm)^W=(\psi^\pm)^W(x,t)$, and $(\psi^\pm_{R,L})^W=(\psi^\pm_{R,L})^W(x,t)$. It is evident now, that in general these two cases are different from physical, mathematical, and computational points of view. In this manner, if we consider the quantum mechanical probability density and its normalization, we are forced to relate the Lorentz invariant probability density to the chosen representation
\begin{equation}\label{pd}
  \Omega^{D,W}\equiv(\bar{\psi^\pm})^{D,W}(\psi^\pm)^{D,W},
\end{equation}
\begin{equation}\label{normal}
  \int d^3x \Omega^{D,W}= \mathbf{1}_4.
\end{equation}
Using of (\ref{pdirac}) by elementary calculation one obtains
\begin{equation}
  \Omega^D=\left[\begin{array}{cc}\dfrac{(\bar{\psi^\pm}_R)^D(\psi^\pm_R)^D+(\bar{\psi^\pm}_L)^D(\psi^\pm_L)^D}{2}\mathbf{1}_2&\dfrac{(\bar{\psi^\pm}_R)^D(\psi^\pm_R)^D-(\bar{\psi^\pm}_L)^D(\psi^\pm_L)^D}{2}\mathbf{1}_2\\
  \dfrac{(\bar{\psi^\pm}_R)^D(\psi^\pm_R)^D-(\bar{\psi^\pm}_L)^D(\psi^\pm_L)^D}{2}\mathbf{1}_2&\dfrac{(\bar{\psi^\pm}_R)^D(\psi^\pm_R)^D+(\bar{\psi^\pm}_L)^D(\psi^\pm_L)^D}{2}\mathbf{1}_2\end{array}\right],
\end{equation}
By application of (\ref{pweyl}) the probability density (\ref{pd}) becomes
\begin{equation}
  \Omega^W=\left[\begin{array}{cc}(\bar{\psi^\pm}_R)^W(\psi^\pm_R)^W\mathbf{1}_2&\mathbf{0}_2\\
  \mathbf{0}_2&(\bar{\psi^\pm}_L)^W(\psi^\pm_L)^W\mathbf{1}_2\end{array}\right].
\end{equation}
Employing the normalization condition (\ref{normal}) in the Dirac representation one obtains the system of equations
\begin{equation}
\dfrac{1}{2}\left(\int d^3x(\bar{\psi^\pm}_R)^D(\psi^\pm_R)^D+\int d^3x(\bar{\psi^\pm}_L)^D(\psi^\pm_L)^D\right)=1,
\end{equation}
\begin{equation}
\dfrac{1}{2}\left(\int d^3x(\bar{\psi^\pm}_R)^D(\psi^\pm_R)^D-\int d^3x(\bar{\psi^\pm}_L)^D(\psi^\pm_L)^D\right)=0,
\end{equation}
which leads to
\begin{eqnarray}
  \int d^3x(\bar{\psi^\pm}_R)^D(\psi^\pm_R)^D&=&1,\label{dirnorm1}\\
  \int d^3x(\bar{\psi^\pm}_L)^D(\psi^\pm_L)^D&=&1.\label{dirnorm2}
\end{eqnarray}
In the case of Weyl representation one receives straightforwardly
\begin{eqnarray}
  \int d^3x(\bar{\psi^\pm}_R)^W(\psi^\pm_R)^W&=&1,\label{weylnorm1}\\
  \int d^3x(\bar{\psi^\pm}_L)^W(\psi^\pm_L)^W&=&1.\label{weylnorm2}
\end{eqnarray}
In this manner one sees that in fact both the conditions (\ref{dirnorm1}), (\ref{dirnorm2}) and (\ref{weylnorm1}), (\ref{weylnorm2})) are invariant with respect to choice of gamma matrices representations
\begin{equation}\label{norma}
  \int d^3x(\bar{\psi^\pm}_{R,L})^{D,W}(\psi^\pm_{R,L})^{D,W}=1,
\end{equation}
that means they are physical. Using of the fact that full space-time evolution is determined as
\begin{eqnarray}
  &&(\psi^\pm_{R,L})^{D,W}(x,t)=U^{D,W}(t,t_0)(\psi^\pm_{R,L})^{D,W}(x,t_0),\\
  &&\left[U^{D,W}(t,t_0)\right]^\dagger U^{D,W}(t,t_0)=\mathbf{1}_2,
\end{eqnarray}
one finds easily the condition
\begin{equation}\label{norma1}
  \int d^3x(\bar{\psi^\pm}_{R,L})^{D,W}(x,t_0)(\psi^\pm_{R,L})^{D,W}(x,t_0)=1.
\end{equation}
By using of the spatial evolution (\ref{psix1}) one obtains the relation
\begin{equation}\label{norma1}
  C\int d^3x\left(\mathbf{1}_2+\dfrac{(x-x_0)_i}{|x-x_0|}\Im\sigma^i\sin\left|2\dfrac{{p_{R,L}^\pm}^0}{\hbar}(x-x_0)_i\right|\right)=1,
\end{equation}
where $C\equiv \left|(\psi^\pm_{R,L})^{D,W}(x_0,t_0)\right|^2$ is a constant, and $\Im{\sigma^i}=\dfrac{\sigma^i-\sigma^{i\dagger}}{2i}$ is a imaginary part of the vector $\sigma^i$. The decomposition $\sigma_i=[\sigma_x,0,\sigma_z]+i[0,-i\sigma_y,0]$ yields $\Im\sigma^i=[0,-i\sigma_y,0]$, and the equation (\ref{norma1}) becomes
\begin{equation}\label{norma2}
  C\int d^3x\left(\mathbf{1}_2-i\dfrac{(x-x_0)_y}{|x-x_0|}\sigma_y\sin\left|2\dfrac{{p_{R,L}^\pm}^0}{\hbar}(x-x_0)_i\right|\right)=1.
\end{equation}
Introducing the change of variables $(x-x_0)_i\rightarrow {x'}_i$ in the following way
\begin{equation}
  {x'}_i\equiv 2\dfrac{{p_{R,L}^\pm}^0}{\hbar}(x-x_0)_i,
\end{equation}
and the effective volume $V'$ due to the vector ${x'}_i$
\begin{equation}\label{volum}
  V'\mathbf{1}_2=\int d^3x'\left\{\mathbf{1}_2-i\sigma_y{x'}_y\dfrac{\sin|x'|}{|x'|}\right\},
\end{equation}
the equation (\ref{norma2}) can be rewritten in the form
\begin{equation}\label{norma2a}
  \int d^3x'V'\mathbf{1}_2=\dfrac{1}{C}\mathbf{1}_2,
\end{equation}
so that one obtains easily
\begin{equation}
  (\psi^\pm_{R,L})^{D,W}(x_0,t_0)=\left(2\dfrac{{p_{R,L}^\pm}^0}{\hbar}\right)^{3/2}\dfrac{1}{\sqrt{V'}}\exp{i\theta_{\pm}},
\end{equation}
where $\theta_{\pm}$ are arbitrary constant phases. The volume (\ref{volum}) differs from the standard one by the presence of the extra axial (y) volume $V_y$
\begin{equation}
  V_y=-i\sigma_y\int d^3x' {x'}_y\dfrac{\sin|x'|}{|x'|},
\end{equation}
which is the axial effect and has nontrivial feature, namely
\begin{equation}
  V_y=\left\{\begin{array}{cc}0&\mathrm{on}~\mathrm{finite}~\mathrm{symmetrical}~\mathrm{spaces}\\
  \infty&\mathrm{on}~\mathrm{infinite}~\mathrm{symmetrical}~\mathrm{spaces}\\
  <\infty&\mathrm{on}~\mathrm{sections}~\mathrm{of}~\mathrm{symmetrical}~\mathrm{spaces}\end{array}\right. .
\end{equation}

One sees now that the normalization is strictly speaking dependent on the choice of a region of integrability. For infinite symmetric spatial regions this procedure is not well defined, because the axial effect is infinite. However, one can consider some reasonable cases that consider the quantum theory on finite symmetric spatial regions. Moreover, the problem of integrability is defined with respect to choice of the initial momentum of the Weyl chiral fields ${p_{R,L}^\pm}^0$. In fact there are many possible nonequivalent physical situations connected with a concrete choice of this eigenvalue. The one of this type situations related to a finite symmetric spatial region, we are going to discuss in the next section as the example of the massive neutrinos model, which in general was solved in this paper.
\section{The reasonable case}
Let us consider finally the reasonable case, that is based on the normalization in a finite symmetrical box and putting by hands the value of initial momenta of the chiral Weyl fields according to the Special Relativity
\begin{equation}\label{choice}
  {p_{R,L}^\pm}^0=\mu_{R,L}^\pm c.
\end{equation}
For that simplified case the normalization discussed in the previous section leads to the following initial data condition
\begin{equation}
  (\psi^\pm_{R,L})^{D,W}(x_0,t_0)=\sqrt{\left(2\dfrac{c}{\hbar}\right)^3\dfrac{\mu^3_{R,L}}{V'}}\exp{i\theta_{\pm}}=\dfrac{1}{\sqrt{V}}\exp{i\theta_{\pm}},
\end{equation}
where $V=\int d^3x$. Introducing the function $E^D(x,y)$
\begin{equation}
  E^D(x,y)\equiv c^2\sqrt{\strut{\left(\dfrac{x+y}{2}\right)^2}+x^2},
\end{equation}
the right- and the left-hand chiral Weyl fields in the Dirac representation take the following form
\begin{eqnarray}
  &&(\psi^\pm_R)^D(x,t)=\Bigg\{\Bigg[\cos\left[\dfrac{t-t_0}{\hbar}E^D(\mu_R^{\pm},\mu_L^{\pm})\right]-\nonumber\\
  &-&i\dfrac{\mu_{\pm}^Dc^2}{E^D(\mu_R^{\pm},\mu_L^{\pm})}\sin\left[\dfrac{t-t_0}{\hbar}E^D(\mu_R^{\pm},\mu_L^{\pm})\right]\Bigg]\exp\left\{-\dfrac{ic}{\hbar}\mu_R^{\pm}(x-x_0)_i\sigma^i\right\}-\nonumber\\
  &-&i\dfrac{\mu_L^{\pm}c^2}{E^D(\mu_L^{\pm},\mu_R^{\pm})}\sin\left[\dfrac{t-t_0}{\hbar}E^D(\mu_L^{\pm},\mu_R^{\pm})\right]\exp\left\{-\dfrac{ic}{\hbar}\mu_L^{\pm}(x-x_0)_i\sigma^i\right\}\Bigg\}\times\nonumber\\
  &\times&\dfrac{1}{\sqrt{V}}\exp\left\{i\left[\theta_{\pm}-\dfrac{(\mu_R^{\pm}-\mu_L^{\pm})c^2}{2\hbar}(t-t_0)]\right]\right\}.
\end{eqnarray}\
and
\begin{eqnarray}
&&(\psi^\pm_L)^D(x,t)=\Bigg\{\Bigg[\cos\left[\dfrac{t-t_0}{\hbar}E^D(\mu_L^{\pm},\mu_R^{\pm})\right]+\nonumber\\
    &+&i\dfrac{\mu_{\pm}^Dc^2}{E^D(\mu_L^{\pm},\mu_R^{\pm})}\sin\left[\dfrac{t-t_0}{\hbar}E^D(\mu_L^{\pm},\mu_R^{\pm})\right]\Bigg]\exp\left\{-\dfrac{ic}{\hbar}\mu_{L}^\pm(x-x_0)_i\sigma^i\right\}-\nonumber\\
  &-&i\dfrac{\mu_R^{\pm}c^2}{E^D(\mu_R^{\pm},\mu_L^{\pm})}\sin\left[\dfrac{t-t_0}{\hbar}E^D(\mu_R^{\pm},\mu_L^{\pm})\right]\exp\left\{-\dfrac{ic}{\hbar}\mu_{R}^\pm(x-x_0)_i\sigma^i\right\}\Bigg\}\times\nonumber\\
  &\times&\dfrac{1}{\sqrt{V}}\exp\left\{i\left[\theta_{\pm}-\dfrac{(\mu_R^{\pm}-\mu_L^{\pm})c^2}{2\hbar}(t-t_0)]\right]\right\}.
\end{eqnarray}
Similarly, introducing the function $E^W(x,y)$
\begin{equation}
  E^W(x,y)\equiv c^2\sqrt{xy+x^2},
\end{equation}
for the studied case the right- and the left-hand chiral Weyl fields in the Weyl representation have a form
\begin{eqnarray}
&&(\psi^\pm_R)^W(x,t)=\dfrac{\exp{i\theta_{\pm}}}{\sqrt{V}}\Bigg\{\Bigg[\cos\left[\dfrac{t-t_0}{\hbar}E^W(\mu_R^{\pm},\mu_L^{\pm})\right]-\nonumber~~~~~~\\
&-&\dfrac{i\mu_R^{\pm}c^2}{E^W(\mu_R^{\pm},\mu_L^{\pm})}\sin\left[\dfrac{t-t_0}{\hbar}E^W(\mu_R^{\pm},\mu_L^{\pm})\right]\Bigg]\exp\left\{-\dfrac{ic}{\hbar}\mu_{R}^\pm(x-x_0)_i\sigma^i\right\}+\nonumber~~~~~~\\
&+&\dfrac{i\mu_L^{\pm}c^2}{E^W(\mu_L^{\pm},\mu_R^{\pm})}\sin\left[\dfrac{t-t_0}{\hbar}E^W(\mu_L^{\pm},\mu_R^{\pm})\right]\exp\left\{-\dfrac{ic}{\hbar}\mu_{L}^\pm(x-x_0)_i\sigma^i\right\}\Bigg\},~~~~~~
\end{eqnarray}
and
\begin{eqnarray}
&&(\psi^\pm_L)^W(x,t)=\dfrac{\exp{i\theta_{\pm}}}{\sqrt{V}}\Bigg\{\Bigg[\cos\left[\dfrac{t-t_0}{\hbar}E^W(\mu_L^{\pm},\mu_R^{\pm})\right]-\nonumber~~~~~~\\
&-&\dfrac{i\mu_L^{\pm}c^2}{E^W(\mu_L^{\pm},\mu_R^{\pm})}\sin\left[\dfrac{t-t_0}{\hbar}E^W(\mu_L^{\pm},\mu_R^{\pm})\right]\Bigg]\exp\left\{-\dfrac{ic}{\hbar}\mu_{L}^\pm(x-x_0)_i\sigma^i\right\}+\nonumber~~~~~~\\
&+&\dfrac{i\mu_R^{\pm}c^2}{E^W(\mu_R^{\pm},\mu_L^{\pm})}\sin\left[\dfrac{t-t_0}{\hbar}E^W(\mu_R^{\pm},\mu_L^{\pm})\right]\exp\left\{-\dfrac{ic}{\hbar}\mu_{R}^\pm(x-x_0)_i\sigma^i\right\}\Bigg\}.~~~~~~
\end{eqnarray}

The ''reasonable case'' considered above is only the example following from the massive neutrinos model given by the massive Weyl equations (\ref{neu}) obtained due to the Snyder model of noncommutative geometry (\ref{nd}), and naturally it is not the only case. Actually there are many other possibilities for determination of the relation between the initial eigenmomentum values ${p_{R,L}^\pm}^0$ and the masses $\mu_{R,L}^{\pm}$ of the right- and left- hand chiral Weyl fields $\psi_{R,L}^\pm$. However, the concrete choice (\ref{choice}) tested in this section presents a crucial reasonability contained in its special-relativistic-like character. Naturally this choice is connected with the special equivalence principle applied to the massive neutrinos in at the beginning of their space-time evolution, \emph{i.e.} $E_{R,L}^\pm=\mu^\pm_{R,L}c^2={p_{R,L}^\pm}^0c$. This case, however, is also nontrivial from the high energy physics point of view \cite{uhep}, namely it is related to the region of ultra-high energies, widely considered in the modern astrophysics (See \emph{e.g.} \cite{uheap} and suitable references therein). So, the presented reasonable case of the massive neutrinos evolution in fact describes their physics in this region, and has possible natural application in ultra-high energy astrophysics.

\section{Outlook}
In this paper we have discussed in some detail the consequences of the massive neutrinos model arising due to the Snyder model of noncommutative geometry. The massive neutrinos model is a consequence of the Dirac equations for a usual relativistic quantum state supplemented by the generalized $\gamma^5$-term \cite{glinka}. In fact, Sidharth has suggested that this term could give a neutrino mass, however, in spite of a good physical intuition he has finished considerations on a laconic statement only, with no any concrete calculations and propositions for a generation mechanism of neutrinos masses \cite{sidharth1}. 

First we have considered the physical status of the Snyder model. By detailed calculation we have shown that, in contrast to Special Relativity theory, within the massive neutrinos model an energy of any original relativistic massive or massless quantum state is strictly renormalized due to a maximal energy, directly related to a minimal scale $\ell$ being the deformation parameter in the Snyder noncommutative geometry. In this manner the Snyder model has received a deep physical sense, that is in some partial relation to the Markov--Kadyshevsky approach \cite{kadyshevsky}.

Next the integrability problem of the massive neutrinos model was detailed discussed. First we have considered the modified Dirac equations, which rewritten in the Schr\"odinger form have yielded manifestly nonhermitian Hamiltonian being a sum of a hermitian and a antihermitian parts. By employing the 4th order approximation of the Zassenhaus formula we have proven that the procedure is not algorithm by the presence of the cyclic problem in exponentialization. Consequently, the Dirac equations are not integrable exactly. By this formal reason we have redefined the integrability problem with respect to the massive Weyl equations corresponding to the Dirac equations. The massive Weyl equations was also rewritten in the Schr\"odinger form, and by using of both the Dirac and the Weyl representations of the Dirac $\gamma$ matrices, we have constructed its analytical exact solutions. We have shown that the normalizability of a solution is correctly defined only for special regions of spatial integration, \emph{i.e.} sections of symmetric spaces or finite symmetric spaces. 

Finally, the case related to ultra-high energy physics and astrophysics was shortly discussed. The all obtained results in general present interesting new physical content. In itself the investigated quantum-mechanical approach to the massive Weyl equation is novel. There are still possible applications of the proposed massive neutrinos model to phenomenology of particle physics and astrophysics, especially in the ultra-high energy region. The open question is the gauge field theory related to the massive neutrinos model, possessing some features of QCD.

\section*{Acknowledgements}

The author benefitted many valuable discussions from Profs. A. B. Arbuzov, I. Ya. Aref'eva, K. A. Bronnikov, I. L. Buchbinder, and V. N. Pervushin. Special thanks are directed to Profs. B. G. Sidharth and S. R. Valluri.

\end{document}